# Self-consistent Capacitance-Voltage Characterization of Gate-all-around Graded Nanowire Transistor

SaeedUzZaman Khan*, Md. Shafayat Hossain, Md. Obaidul Hossen, Fahim Ur Rahman,
Rifat Zaman, and Quazi D. M. Khosru
Department of Electrical and Electronic Engineering, Bangladesh University of Engineering and Technology
Dhaka-1000, Bangladesh
*anim.buet@gmail.com

*Abstract*—This paper presents a self-consistent numerical model for calculating the charge profile and gate capacitance and therefore obtaining C-V characterization for a gate-all-around graded nanowire MOSFET with a high mobility axially graded $In_{0.75}Ga_{0.25}As$ + $In_{0.53}Ga_{0.47}As$ channel incorporating strain and atomic layer deposited $Al_2O_3$/20nm Ti gate. C-V characteristics with introduction and variation of In-composition grading and also grading in doping concentration are explored. Finite element method has been used to solve Poisson's equation and Schrödinger's equation self-consistently considering wave function penetration and other quantum effects to calculate gate capacitance and charge profile for different gate biases. The device parameters are taken from a recently introduced experimental device.

## I. Introduction

III-V MOSFETs are replacing conventional Si MOSFETs to suppress short-channel effects (SCE). III-V FinFETs [1]-[2] and multi-gate quantum-well FETs [3] provide performance improvements of III-V FETs with deep submicron gate lengths. As predicted by International Technology Roadmap for Semiconductors [4], Silicon-On-Insulator (SOI) and multiple gate devices like double gate FinFETs and gate-all-around (GAA) MOSFETs offer good electrostatic control required for gate lengths around 25 nm. GAA structure for Si- CMOS has been proven most resistant to SCE [5]-[7]. Recently, gate-all-around $In_{0.53}Ga_{0.47}As$ MOSFET has been experimentally demonstrated [8] having the shortest channel length ($L_{ch}$=50nm) to date. Graded Nanowire is a recent concept which is yet to be introduced in experimental devices. We propose an In-composition graded channel with improved interface where doping can also be graded in GAA structure. This paper presents numerical C-V characterization, charge profile of the device in [8] changing the channel to an axially graded nanowire $In_{0.75}Ga_{0.25}As$ + $In_{0.53}Ga_{0.47}As$ channel and then the comparison with the original device. QM C-V of the original device has been taken from our previous work [9]. The effects of change in grading and introduction of graded doping on C-V characteristics are also explored.

## II. Device Structure

The device modeled is an axially graded $In_{0.75}Ga_{0.25}As$ + $In_{0.53}Ga_{0.47}As$ nanowire FET. It is a gate-all-around device comprising four gates connected together on all four sides of an oxide ($Al_2O_3$) - axially graded nanowire $In_{0.75}Ga_{0.25}As$ + $In_{0.53}Ga_{0.47}As$ structure having square-shaped cross-section. The channel material is axially graded $In_{0.75}Ga_{0.25}As$ + $In_{0.53}Ga_{0.47}As$ and 10 nm thick $Al_2O_3$ serves as the gate dielectric. The channel material adjacent to $Al_2O_3$ is $In_{0.75}Ga_{0.25}As$ because, $Al_2O_3$- $In_{0.75}Ga_{0.25}As$ interface exhibits less interface trap charge [10] due to the reduction of the density of the $Ga^{3+}$ oxidation state with increasing In concentration, as the gallium concentration is concomitantly reduced leading to smaller number of defect states at the interface [11]. At the same time, percentage of donor-like traps increases in the $D_{it}$ profile which is explained by the charge-neutrality-level model for III-VMOSFETs [12].

$In_{0.75}Ga_{0.25}As$ is 5nm wide on each side i.e. comprising 10nm width and $In_{0.53}Ga_{0.47}As$ is 20nm wide. The source and drain regions are formed with n+ $In_{0.75}Ga_{0.25}As$ + $In_{0.53}Ga_{0.47}As$. The substrate is 30nm×30nm p+ InP. The doping for p- channel is Na=$2 \times 10^{16}$ cm$^{-3}$. Ti is used as gate metal for adjusting flat-band with the device in [8] for better performance-comparison.

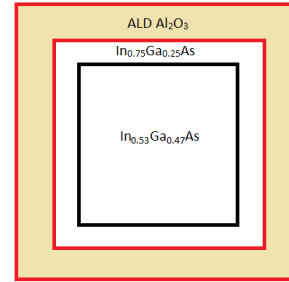

Fig.1. Cross section of an inversion-mode GAA n-channel 10nm $In_{0.75}Ga_{0.25}As$ ($2\times10^{16}$/cm$^3$) + 20nm $In_{0.75}Ga_{0.25}As$ ($2\times10^{16}$/cm$^3$) MOSFET with ALD10nm $Al_2O_3$/20nm Ti gate.

## III. Simulator Development

The developed self-consistent simulator involves numerical solution of two partial differential equations in a coupled manner, namely Poisson's equation and Schrödinger's equation along with numerical integration. Finite Element Method (FEM) is used to solve the PDE's. COMSOL Multiphysics [13] is used as the PDE solver tool which is invoked from MATLAB [14] by using the scripting language of COMSOL. Poisson's equation

$$-\nabla . \ (\varepsilon \ \nabla V) = \rho \quad (1)$$

At the external boundaries of the gates *Dirichlet* i.e. fixed voltage boundary condition is used. *Neumann* i.e. continuous electric flux boundary condition is used at all internal boundaries.

Schrödinger's equation as given by the effective mass approximation

$$\left(-\frac{\hbar^2}{2m_{ds}^*}\nabla^2 - qV(x,y)\right)\psi_j(x,y) = E_j\psi_j(x,y) \quad (2)$$

$m_{ds}^*$ is the density of states effective mass and $E_j$ and $\psi_j$ are the minimum energy and corresponding wave function of the $j^{th}$ sub-band respectively. For Schrödinger's equation, all boundaries are kept as open boundaries to allow wave function penetration.

*A. Self-consistent Modelling*

In case of self-consistent modeling considering wave function penetration and other quantum effects, Poisson's equation is solved in 2-D self-consistently with 2-D Schrödinger's equation [15].

In the oxide region, charge density ($\rho$) is zero. For channel region, charge density is obtained according to the following expression

$$\rho(y,z) = q\sum_i \int_{E_i}^{\infty} D(E)\,f(E)\,|\psi_i(y,z)|^2 dE \quad (3)$$

Where $i$ is the number of sub-band, $D(E)$ is 1-D density of states and $f(E)$ is Fermi-Dirac distribution function with respect to source/drain Fermi level. $E_i$ and $\psi_i$ are Eigen energies and corresponding normalized wave functions obtained from 2-D Schrödinger's equation.

*B. C-V Modelling*

For 2-D cross section, constant potential distribution along z-direction is assumed. Gate capacitance per unit channel length is

$$C_G = \frac{dQ_{InGaAs}}{dV_G} \quad (4)$$

Where, $Q_{InGaAs}$ is the charge deposited inside InGaAs which is obtained from

$$Q_{InGaAs} = \int_x \int_y \rho(x,y)\,dy\,dx \quad (5)$$

Furthermore, noise-free numerical derivatives were calculated following the algorithm by Savitzky and Golay [16].

*C. Strain Calculation*

The effect of biaxial compressive strain on the channel layer is incorporated in our adapted model taking account of the shifting of conduction and valence band edges along with the change of effective masses. The strain splits the valence band at the zone center and shifts the spin-orbit band. The degenerated HH and LH bands split into higher HH and lower LH respectively. This shifting results in an increase in the effective band gap. The amount of shifting is calculated using the following formulas [17]:

$$E_{hh}(0) = E_v^0 - P_\varepsilon - Q_\varepsilon \quad (6)$$

$$E_{lh}(0) = E_v^0 - P_\varepsilon + \frac{1}{2}\left[Q_\varepsilon - \Delta_0 + \sqrt{\Delta_0^2 + 9Q_\varepsilon^2 + 2Q_\varepsilon\Delta_0}\right] \quad (7)$$

$$E_{SO}(0) = E_v^0 - P_\varepsilon + \frac{1}{2}\left[Q_\varepsilon - \Delta_0 - \sqrt{\Delta_0^2 + 9Q_\varepsilon^2 + 2Q_\varepsilon\Delta_0}\right] \quad (8)$$

$$E_c(0) = E_v(0) + E_g + a_c(\varepsilon_{xx} + \varepsilon_{yy} + \varepsilon_{zz}) \quad (9)$$

$$P_\varepsilon = -a_c(\varepsilon_{xx} + \varepsilon_{yy} + \varepsilon_{zz}) \quad (10)$$

$$Q_\varepsilon = -\frac{b}{2}(\varepsilon_{xx} + \varepsilon_{yy} - 2\varepsilon_{zz}) \quad (11)$$

Here, $\varepsilon_{zz}$ is the relative change of lattice period in the perpendicular direction and $\varepsilon_{xx}$ and $\varepsilon_{yy}$ are the relative change in lattice period in the in-plane direction respectively. $\Delta_0$ is the split-off energy. Factors $a_c$ and $a_v$ are hydrostatic deformation potentials; while b is the shear deformation potential. Biaxial compressive strain causes the curvatures of the energy band structures and consequently effective masses to change. The hole effective masses of the channel and substrate are calculated using the well-known Luttinger parameters $\gamma_1$, $\gamma_2$ and $\gamma_3$ for k=0 [17].

Table I. Hole Effective mass

|  | Normal Mass, $m_z$ | Transverse Mass, $m_t$ |
|---|---|---|
| HH ($m_{hh}/m_0$) | $\dfrac{1}{\gamma_1 - 2\gamma_2}$ | $\dfrac{1}{\gamma_1 + \gamma_2}$ |
| LH ($m_{lh}/m_0$) | $\dfrac{1}{\gamma_1 + 2\gamma_2 f_+}$ | $\dfrac{1}{\gamma_1 - \gamma_2 f_+}$ |
| SO ($m_{so}/m_0$) | $\dfrac{1}{\gamma_1 + 2\gamma_2 f_-}$ | $\dfrac{1}{\gamma_1 - \gamma_2 f_-}$ |

Here $f_\pm$ is the strain factor calculated from the strain parameter, s

$$f_\pm = \frac{2s\left[1+1.5\left(s-1\pm\sqrt{1+2s+9s^2}\right)\right]+6s^2}{0.75\left(s-1\pm\sqrt{1+2s+9s^2}\right)^2 + s-1\pm\sqrt{1+2s+9s^2}-3s^2} \quad (12)$$

s is the strain parameter, $s = \dfrac{Q_E}{\Delta_0} \quad (13)$

## IV. RESULTS AND DISCUSSIONS

Fig. 2 shows the 1st eigen energy variation with gate voltage to indicate the inversion point. Energy band diagram, Probability density of the carriers and charge concentration for both inversion and accumulation region are on Fig. 2-6. Charge profile and voltage drop in both oxide and channel region are on Fig. 7. Fig. 8 shows the QM C-V for the device and the later figures (Fig. 9-11) shows the variation in C-V characteristics with change in In composition, grading configuration and different graded doping configuration.

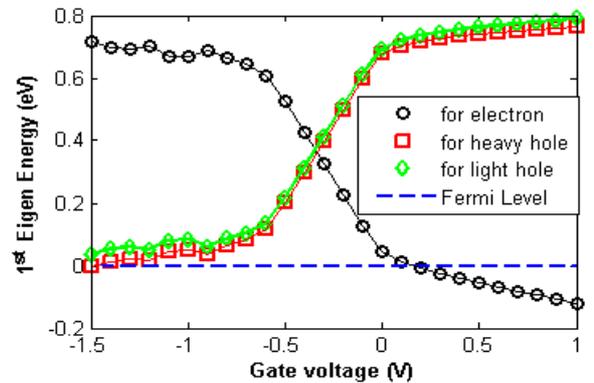

Fig.2. Variation of 1st Eigen energy of electron, heavy-hole and light-hole with gate voltage. It shows that strong inversion starts when 1st Eigen state of electron crosses Fermi level.

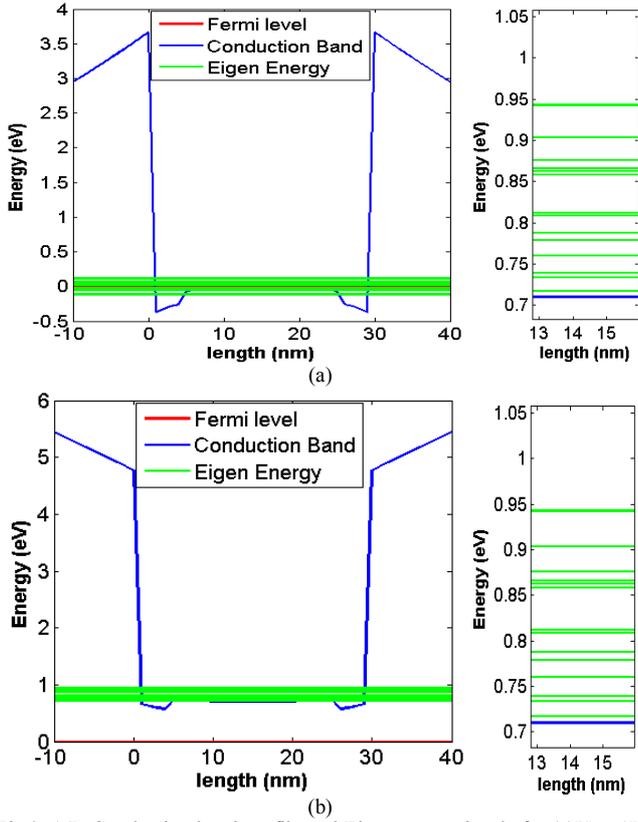

Fig.3. 1-D Conduction band profile and Eigen energy levels for (a)$V_G$ =1V (b) $V_G$ =-1.5V. Energy level splitting is shown in magnified versions.

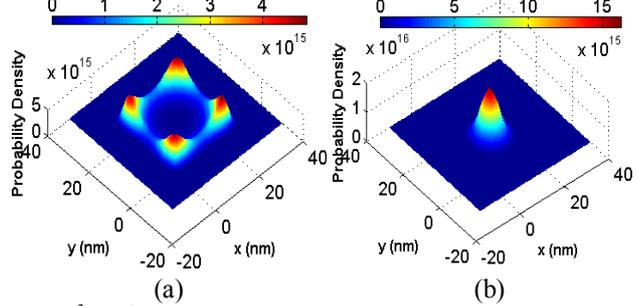

Fig.4. $|\psi|^2$ for 1st Eigen energy for (a) electron & (b) heavy-hole for $V_G$ =1V

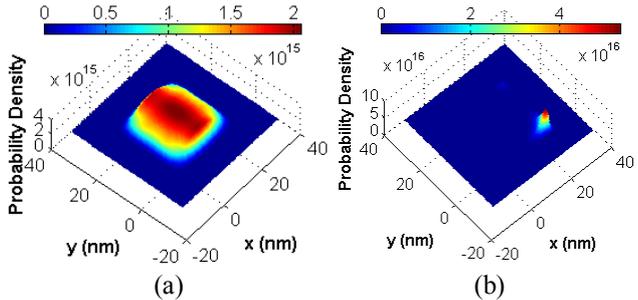

Fig.5. $|\psi|^2$ for 1st Eigen energy for (a) electron & (b) heavy-hole for $V_G$ =-1.5V

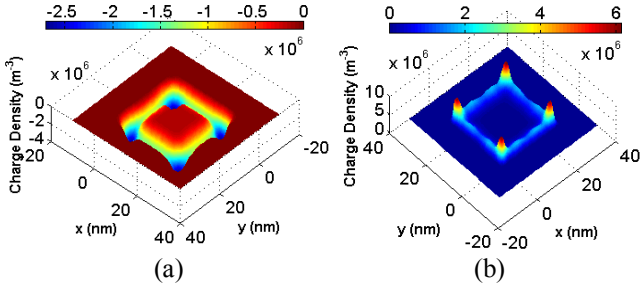

Fig.6. 3-D plot of charge concentration profile for (a) $V_G$ =1V (b) $V_G$ =-1.5V

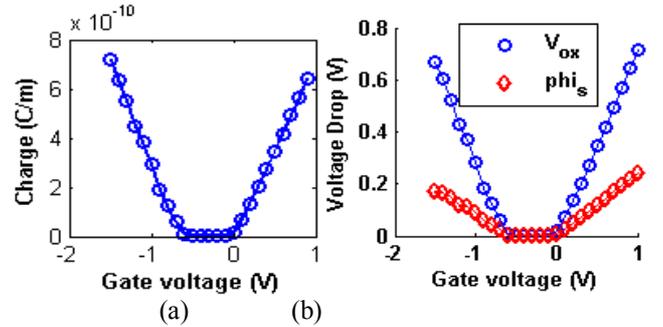

Fig.7. (a) Quantum Q-V per unit channel length. (b)Voltage drop in oxide and semiconductor region showing similar shape as Q-V. A fitting parameter may be described to form Q-V from this curve. Voltage drop is less in semiconductor region than in oxide region.

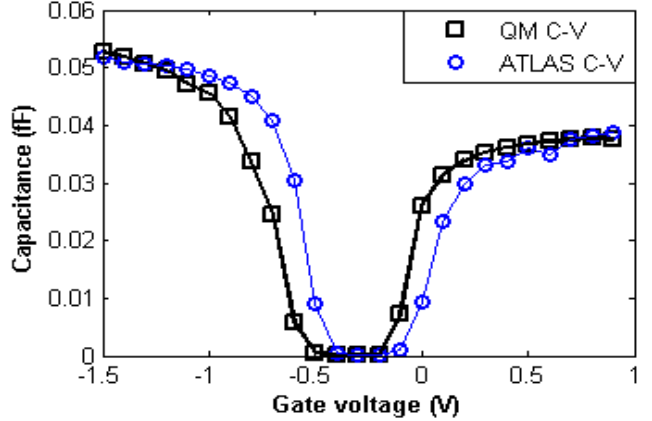

Fig.8. QM C-V. ATLAS simulation is included for comparison. As, $m_e^* < m_h^*$, saturation capacitance is less in inversion region.

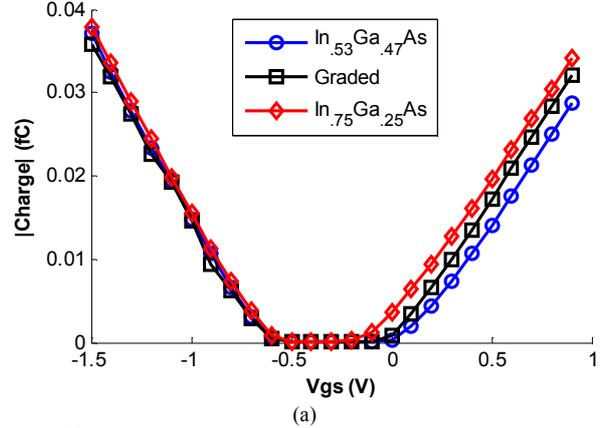

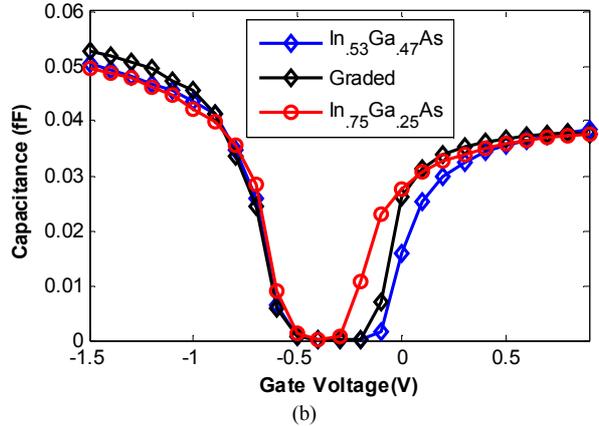

Fig.9.(a) Q-V & (b) C-V comparison of graded nanowire with $In_{0.53}Ga_{0.47}As$ and $In_{0.75}Ga_{0.25}As$ channel GAAFET. Saturation capacitance is almost invariant with channel material. Threshold voltage is in between of $In_{0.75}Ga_{0.25}As$ & $In_{0.53}Ga_{0.47}As$ devices

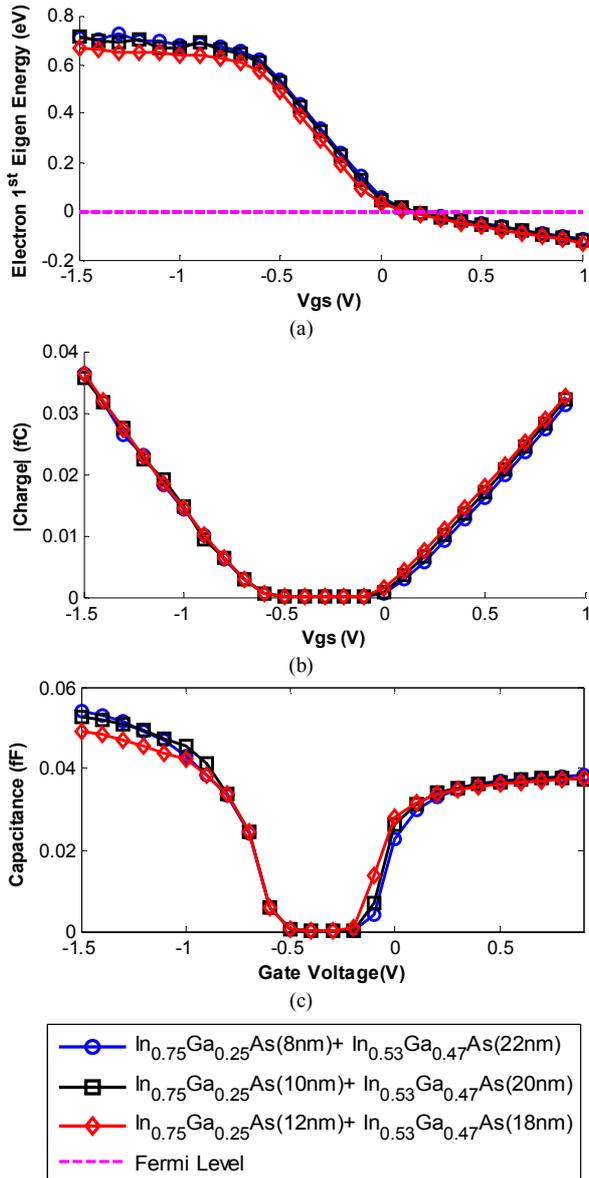

Fig.10. (a) 1st Eigen energy (b) Q-V & (c) C-V variation with change in grading configuration i.e. the comparative width of the InGaAs layers. No change in saturation capacitance. Threshold voltage is shifted to the left with the increased comparative width of $In_{0.75}Ga_{0.25}As$ layer.

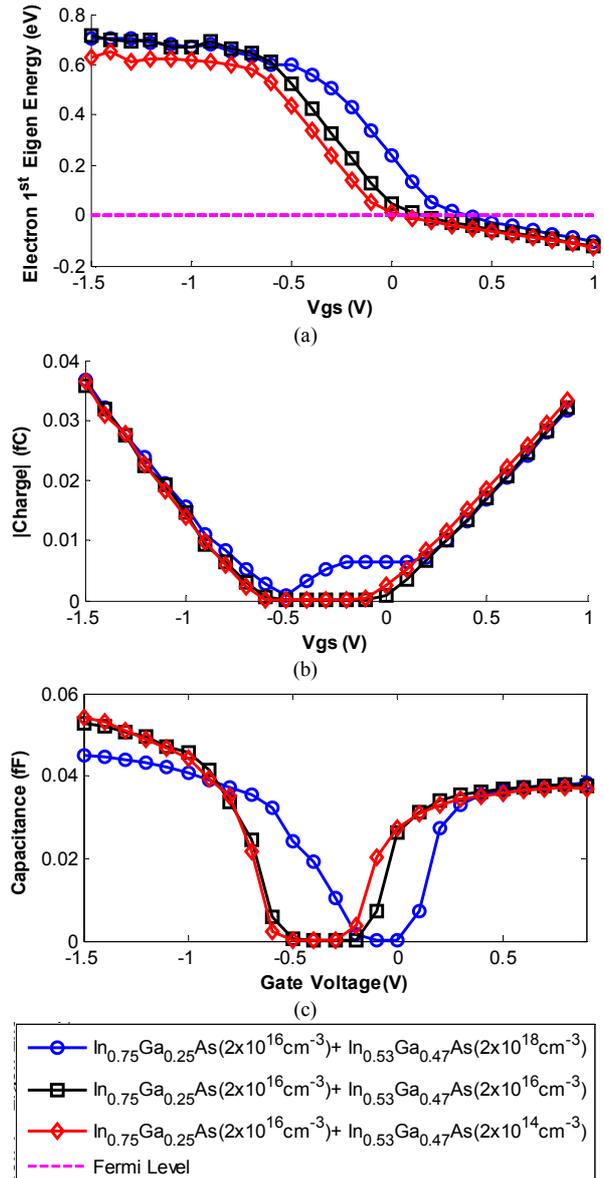

Fig.11. (a) 1st Eigen energy (b) Q-V & (c) C-V variation with change in doping of inner $In_{0.53}Ga_{0.47}As$ layer keeping the outer layer doping fixed. If the inner layer is lightly doped than the outer layer, saturation capacitance remains unchanged and threshold voltage becomes more negative. On the other hand, if the inner layer is heavily doped, saturation capacitance in the inversion region remains the same but that of accumulation region is decreased. Threshold voltage also becomes more positive.

## V. CONCLUSION

This work presents a numerical model for self-consistent Quantum mechanical C-V characterization of axially graded $In_{0.75}Ga_{0.25}As + In_{0.53}Ga_{0.47}As$ nanowire MOSFET which is yet to be explored in experimental devices. The model is developed using finite element method for PDE solving and matched with ATLAS simulation results. C-V variations with change in grading and introduction of graded doping have been explored in detail.